\documentclass[epj]{webofc}
\usepackage[varg]{txfonts}
\usepackage{microtype}
\usepackage{mathtools}
\usepackage{tikz}
\usepackage{pgfplots}
\usepackage{orcidlink}  
\pgfplotsset{compat=1.18}
\hypersetup{colorlinks=true,linkcolor=blue!45!black,
  citecolor=blue!45!black,urlcolor=blue!45!black,
  pdftitle={Searching for the pseudoscalar glueball with fractional determinants},
  pdfauthor={Francesco Giacosa},
  pdfsubject={MESON2026 proceedings: pseudoscalar glueball, axial anomaly and X(2370)}}
\newcommand{\Bf}{\mathcal{B}}
\newcommand{\doiurl}[1]{\href{https://doi.org/#1}{\nolinkurl{#1}}}
\newcommand{\arxivurl}[1]{\href{https://arxiv.org/abs/#1}{arXiv:#1}}
\woctitle{18th International Workshop on Meson Physics (MESON2026)}

\begin{document}

\title{Searching for the pseudoscalar glueball with fractional determinants}
\author{\firstname{Francesco} \lastname{Giacosa \orcidlink{0000-0002-7290-9366}}\inst{1,2}\fnsep
  \thanks{\email{francesco.giacosa@ujk.edu.pl}}}
\institute{Institute of Physics, Jan Kochanowski University, ul. Uniwersytecka 7,
  25-406 Kielce, Poland
  \and
  Institute for Theoretical Physics, Goethe University, Max-von-Laue-Stra\ss e 1,
  60438 Frankfurt am Main, Germany}

\abstract{The pseudoscalar glueball is closely linked to the axial anomaly,
making its decays a natural place to study anomalous meson interactions.
After a brief account of the search and of the BESIII evidence for $X(2370)$,
I discuss the determinant interaction and its recent extension
(arXiv:2610.00484) to fractional powers of the normalized determinant phase.
The emphasis is on the connection between the gluonic vacuum and hadronic
decays, and on the opportunities opened by the new experimental results and
this new class of anomalous interactions.}
\maketitle

\section{Brief recall of the pseudoscalar glueball}
\label{sec:intro}

Gluons interact with each other. This simple statement, which distinguishes
QCD from an Abelian gauge theory, suggests an equally simple question: can
gluons form hadrons without valence quarks? Glueballs are the answer expected
from the nonperturbative dynamics of QCD. Finding them, however, has proved much harder than anticipating their
existence; for recent overviews, see Refs.~\cite{Morningstar,ChevalierMathieu}.

In Yang--Mills (YM) theory, glueballs form the physical spectrum of the
pure gauge theory and, as such, have been established in various lattice
calculations~\cite{Chen,AT} (see also the comparison in Ref.~\cite{Trotti}).
In QCD with dynamical quarks, they can mix with quark--antiquark and
multihadron configurations. Experimentally, the aim is therefore to identify
a predominantly gluonic resonance, rather than a perfectly pure glueball.
Neither a mass match nor a single favorable decay mode can settle this
issue.

The pseudoscalar sector has its own long history. The resonances once grouped
under $\eta(1440)$, and in particular $\eta(1405)$ and $\eta(1475)$ \cite{pdg}, attracted considerable
attention as glueball candidates; see the review in Ref. \cite{Masoni}. Chiral mixing schemes involving a pseudoscalar glueball
and radially excited $\eta$ and $\eta'$ mesons explored this
possibility~\cite{Gutsche}. Modern pure-gauge lattice calculations instead
place the lightest $0^{-+}$ glueball near $2.5$--$2.6$~GeV, well above the
scalar ground state~\cite{Chen,AT}. Hence, a predominantly gluonic state around
$1.4$~GeV is difficult to reconcile with that spectrum. This does not forbid a gluonic admixture in a lighter
meson, but it changes where one should look for the dominant glueball state.
In addition,  the pseudoscalar glueball is also peculiar because it is linked to the axial anomaly \cite{tHooft,rosenzweig}, which represents a fundamental nonperturbative feature of QCD.  

The recent BESIII measurements (see below) make this discussion timely and allow a more ambitious question: can the decays of a glueball
candidate tell us something about the axial anomaly itself? This is the
perspective developed in Ref. \cite{Fractional}, which provides the main thread of this contribution.

\section{What do we know about \texorpdfstring{$X(2370)$}{X(2370)}?}
\label{sec:bes}

BESIII first observed $X(2370)$ in $J/\psi\to\gamma\pi^+\pi^-\eta'$ in
2011~\cite{BES2011}. The partial-wave analysis of
$J/\psi\to\gamma K_S^0K_S^0\eta'$ subsequently established
$J^{PC}=0^{-+}$~\cite{BES2024}. Further observations in
$K_S^0K_S^0\pi^0$ and $\pi^0\pi^0\eta$ have enlarged the available decay
information. The combined resonance parameters reported in the latter
study are $M_X=2359^{+13}_{-14}$~MeV and
$\Gamma_X=170^{+44}_{-29}$~MeV~\cite{BES2026modes}.

The recent BESIII study brings together several arguments for a predominantly
gluonic interpretation: sizable radiative $J/\psi$ production, several
three-pseudoscalar decay modes, and suppression of $K^*(892)\bar K$ and of the
radiative channels $\gamma\omega$ and $\gamma\phi$~\cite{BES2026glue}.
These are important pieces of evidence. Their interpretation still involves
assumptions about flavor symmetry, phase space and competing quark-model
assignments; a flavor-singlet pattern alone is not a measurement of the
glueball fraction.

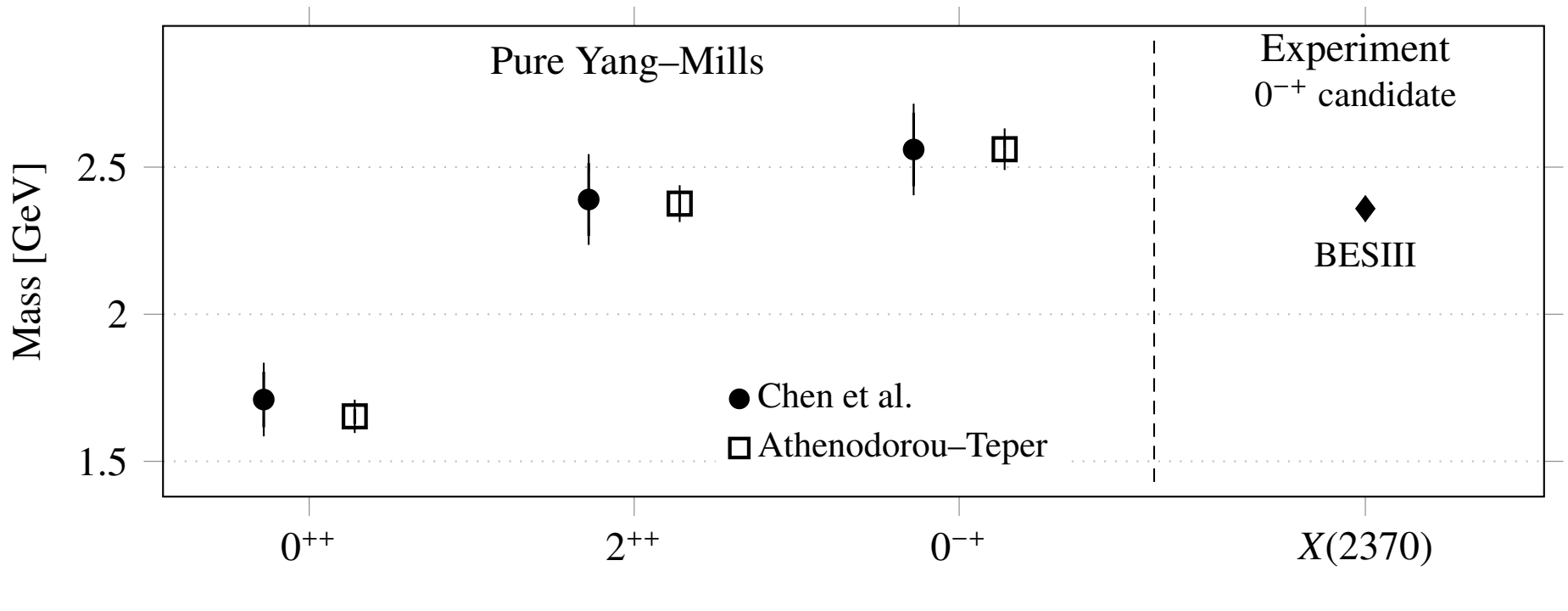
\begin{figure}[!b]
\centering
\begin{tikzpicture}
\begin{axis}[
  width=0.95\linewidth,height=5.25cm,
  xmin=0.55,xmax=4.80,ymin=1.38,ymax=2.98,
  xtick={1,2,3,4.25},
  xticklabels={$0^{++}$,$2^{++}$,$0^{-+}$,{$X(2370)$}},
  ytick={1.5,2.0,2.5},
  ylabel={Mass [GeV]},
  tick label style={font=\small},label style={font=\small},
  ymajorgrids=true,grid style={dotted},
  tick align=outside,
  legend style={font=\footnotesize,draw=none,
    at={(0.40,0.03)},anchor=south west,cells={anchor=west}},
  legend image post style={mark size=2.1pt},
  clip=false]
\addplot[only marks,mark=*,mark size=2.2pt,
  error bars/.cd,y dir=both,y explicit,
  error bar style={line width=0.6pt},error mark options={mark size=2pt}]
  coordinates {(0.86,1.710) +- (0,0.09434)
               (1.86,2.390) +- (0,0.12369)
               (2.86,2.560) +- (0,0.12500)};
\addlegendentry{Chen et al.}
\addplot[only marks,mark=square,mark size=2.5pt,
  mark options={fill=white,line width=0.8pt},
  error bars/.cd,y dir=both,y explicit,
  error bar style={line width=0.6pt},error mark options={mark size=2pt}]
  coordinates {(1.14,1.653) +- (0,0.026)
               (2.14,2.376) +- (0,0.032)
               (3.14,2.561) +- (0,0.040)};
\addlegendentry{Athenodorou--Teper}
\addplot[only marks,mark=diamond*,mark size=2.7pt,forget plot,
  error bars/.cd,y dir=both,y explicit,
  error bar style={line width=0.6pt},error mark options={mark size=2pt}]
  coordinates {(4.25,2.359) += (0,0.013) -= (0,0.014)};
\draw[densely dashed] (axis cs:3.60,1.43) -- (axis cs:3.60,2.93);
\node[font=\small] at (axis cs:1.98,2.85) {Pure Yang--Mills};
\node[font=\small,align=center] at (axis cs:4.22,2.83)
  {Experiment\\[-1pt]\footnotesize $0^{-+}$ candidate};
\node[font=\footnotesize,anchor=north] at (axis cs:4.25,2.30)
  {BESIII};
\end{axis}
\end{tikzpicture}
\caption{The lowest scalar, tensor and pseudoscalar glueball masses from
Chen et al.~\cite{Chen} and Athenodorou--Teper~\cite{AT}, using their
respective scale settings as collected in Ref.~\cite{Trotti}.
Chen et al.'s two quoted errors are combined in quadrature; the
Athenodorou--Teper errors shown are statistical.
The separate experimental column shows the $X(2370)$ mass~\cite{BES2026modes},
not its width.}
\label{fig:spectrum}
\end{figure}

Figure~\ref{fig:spectrum} places the candidate alongside representative
lattice results. The comparison gives useful orientation, but it is not an
identification: the lattice points belong to pure Yang--Mills theory, whereas
$X(2370)$ is a resonance in full QCD. The mass motivates the search; the
detailed production and decay properties must do the rest.

For testing decay models, the measured products
$\Bf(J/\psi\to\gamma X)\Bf(X\to f)$ are particularly useful.
In ratios of two such products, the common radiative production factor
cancels. One can therefore compare relative partial widths without knowing
the absolute production rate of $X(2370)$.
The channels $K\bar K\pi$, $\eta\pi\pi$, $\eta'\pi\pi$ and
$\eta'K\bar K$ now provide such a test~\cite{BES2026glue}.

\section{The determinant: a natural starting point}
\label{sec:det}

The connection to the anomaly is direct. The gluonic topological charge
density has $J^{PC}=0^{-+}$ and enters the divergence of the flavor-singlet
axial current. It can therefore interpolate a pseudoscalar glueball.
At the hadronic level, scalar and pseudoscalar mesons are collected in
$\Phi=\mathcal{S}+i\mathcal{P}$, with
$\Phi\to U_L\Phi U_R^\dagger$ \cite{eLSMreview}. The determinant is invariant under
$SU(3)_L\times SU(3)_R$, but not under the axial $U(1)_A$ transformation.
The familiar 't Hooft interaction \cite{tHooft,tHooft2}
\begin{equation}
 \mathcal{L}_A=-\xi_1\left(\det\Phi+\det\Phi^\dagger\right)
 \label{eq:anomaly}
\end{equation}
realizes this anomalous symmetry breaking in a mesonic theory. Its
instanton origin and extensions to other mesons have been investigated in
Refs.~\cite{Mixing2018,Anomaly2024}.

A corresponding pseudoscalar-glueball field $P$ couples to mesons as:
\begin{equation}
\mathcal{L}_{P\Phi}^{\rm det}
 =i c_{P\Phi}P
   \left(\det\Phi-\det\Phi^\dagger\right).
 \label{eq:ordinary}
\end{equation}
This Hermitian interaction preserves parity and the chiral
symmetry, but not the axial $U(1)_A$ symmetry.
Ref. ~\cite{Eshraim} studied this interaction in
2013 within the extended Linear Sigma Model
(eLSM), whose vacuum phenomenology and broader applications are described in
Refs.~\cite{eLSMreview,Parganlija}.

The calculations in Ref. \cite{Eshraim} found
a prominent $K\bar K\pi$ channel, nonzero $\eta^{(\prime)}\pi\pi$ channels,
and no three-pion decay in the isospin-symmetric treatment. It also included
scalar--pseudoscalar final states, whose subsequent decays can feed the same
three-body channels. Later, Ref.~\cite{Anomaly2024} related the anomalous coupling to a dilute-instanton-gas estimate,
permitting estimates of absolute partial widths.

The determinant is thus well motivated, but its simplest form need not be
the whole story. Higher anomalous couplings can have important consequences,
including for the chiral phase transition~\cite{Columbia}.
Another extension, the polydeterminant, combines different mesonic matrices
and is useful for interactions involving excited multiplets and nonzero
spin~\cite{Polydet}. 

Very recently, fractional determinant powers address a different
question: the structure of the anomalous interaction inherited from the
$\theta$ dependence of the gluonic vacuum~\cite{Fractional,Pisarski2026}.

\section{From the \texorpdfstring{$\theta$}{theta} vacuum to fractional determinants}
\label{sec:fractional}

The central idea of Ref.~\cite{Fractional} is to start with the gluonic
vacuum; see Refs. \cite{sannino,bonannorev} for recent reviews.  The scalar glueball describes the trace anomaly through a dilaton
potential, while the pseudoscalar glueball couples to the CP-odd part of
the $\theta$ dependence.  The interactions are expanded in harmonics of
$\theta/N_c$, the relevant large-$N_c$ variable on a given vacuum branch.
This also connects the $\theta$ dependence of the vacuum energy, condensate
and glueball masses~\cite{Fractional}. Remarkably, good agreement with lattice YM results of Refs. \cite{bonati,bonanno} is achieved.

The mesonic fields enter through the anomalous combination (e.g. Refs. \cite{witten,DiVecchia})
\begin{equation}
 \Theta=\theta-i\ln U_\Phi,
 \qquad U_\Phi\equiv\frac{\det\Phi}{|\det\Phi|}.
 \label{eq:Theta}
\end{equation}
Thus it is the phase of the determinant that is combined with $\theta$.
Replacing $\theta$ by $\Theta$ in the gluonic interactions, and setting the
scalar glueball to its vacuum value, generates couplings of the form
\nopagebreak
\begin{equation}
 \mathcal{L}_{P\Phi}^{\rm phase}
 =iP\sum_{n\geq1}b_n
 \left[e^{in\theta/N_c}U_\Phi^{\,n/N_c}
       -e^{-in\theta/N_c}(U_\Phi^*)^{n/N_c}\right].
 \label{eq:tower}
\end{equation}
The real coefficients $b_n$ contain the gluonic couplings and vacuum
scales. For three colors, the first determinant powers are $1/3$, $2/3$
and $1$. Fractional powers are therefore not inserted merely as a
phenomenological alternative: they emerge from the way the anomalous
mesonic phase enters the gluonic vacuum~\cite{Fractional}.

Two comments are important. First, a single branch depending on
$\theta/N_c$ is not generally $2\pi$ periodic. The physical vacuum energy
is obtained by selecting the lowest of the branches related by
$\theta\to\theta+2\pi k$; a $2\pi$ shift permutes these branches.
Second, the modulus $|\det\Phi|$ in Eq.~\eqref{eq:Theta} depends on the
meson fields. It is part of the interaction, not a constant that can simply
be absorbed into the coupling. Keeping this dependence changes the decay
vertices compared with a frozen-modulus approximation~\cite{Fractional}.

In the flavor-symmetric limit, the vacuum matrix is proportional to the
identity, $\Phi_0=v\,\mathbf{1}$. Setting aside scalar fluctuations, one
writes $\Phi=\Phi_0+i\mathcal{P}$ and expands about this nonzero vacuum to
obtain the glueball-meson vertices. Flavor breaking and the detailed vertices are treated in
Ref.~\cite{Fractional}; implications for the chiral phase transition are
discussed in Ref.~\cite{Pisarski2026}.

\section{What do decays tell us?}
\label{sec:decays}

The recent experimental progress makes the anomalous interaction testable.
Within the direct three-pseudoscalar decay approximation, the normalized
determinant phase leads to a significantly improved description of the
available $X(2370)$ decay data as compared to the ordinary determinant
vertex. 
\begin{table}[!t]
\centering
\small
\setlength{\tabcolsep}{6pt}
\caption{Ratios of the measured and predicted three-pseudoscalar decay
channels, normalized to $K\bar K\pi$. The experimental ratios are obtained
from the BESIII product branching fractions~\cite{BES2026glue}. The fractional-phase predictions are
from Ref.~\cite{Fractional}; the ordinary determinant results are from
Ref.~\cite{Eshraim}.}
\label{tab:decayratios}
\begin{tabular}{lccc}
\hline
Ratio &
Experiment &
Fractional phase &
Ordinary determinant \\
\hline
$\Gamma_{\eta\pi\pi}/\Gamma_{K\bar K\pi}$
& $0.98^{+0.37}_{-0.39}$ & $1.34$ & $0.36$ \\

$\Gamma_{\eta'\pi\pi}/\Gamma_{K\bar K\pi}$
& $0.60^{+0.18}_{-0.31}$ & $0.69$ & $0.19$ \\

$\Gamma_{\eta'K\bar K}/\Gamma_{K\bar K\pi}$
& $0.120^{+0.045}_{-0.044}$ & $0.096$ & $0.023$ \\
\hline
\end{tabular}
\end{table}
The ratios in Table~\ref{tab:decayratios} are evaluated at $\theta=0$, using
$M_X=2.359$~GeV, the physical isospin-averaged pseudoscalar masses,
the mixing angle $\varphi=-44.6^\circ$, and
$f_\pi=92.2$~MeV, $f_K=110$~MeV. (Thus, flavor-symmetry breaking is retained.)
The full field-dependent factor $|\det\Phi|$ is kept throughout.
Since only ratios are considered, the common production factor and overall
coupling normalization cancel.

The direct tree-level amplitudes are momentum independent and are integrated
over the physical three-body phase space, including the appropriate charge
multiplicities and identical-particle factors~\cite{Fractional}. The resulting
ratios are collected in Table~\ref{tab:decayratios}. For the corresponding
fit to the four charge-summed product branching fractions, statistical and
systematic uncertainties are combined in quadrature and the experimental
inputs are treated as uncorrelated, since no covariance matrix is available.

Fitting one common normalization gives
\[
\text{ordinary determinant (Eq.~(\ref{eq:ordinary})):}\qquad
\chi^2/\mathrm{d.o.f.}=13.2/3 ,
\]
whereas for the normalized determinant phase one finds
\[
\text{normalized determinant phase (Eq.~(\ref{eq:tower})):}\qquad
\chi^2/\mathrm{d.o.f.}=1.82/3 .
\]
Thus, the fractional interaction provides a substantially better description
of the available decay pattern.

The harmonic universality has a simple origin. At cubic order, the normalized
determinant phase contains a single-trace and a triple-trace contribution.
The latter involves only the isoscalar fields $\eta_N$ and $\eta_S$ and hence
does not contribute to the four channels entering Table~\ref{tab:decayratios}.
Consequently, all four direct amplitudes are proportional to the same
harmonic-dependent combination
\[
\sum_{n\geq1} n b_n ,
\]
which cancels from their ratios. Every individual harmonic, as well as any
combination for which this coefficient is nonzero, therefore predicts the
same ratios~\cite{Fractional}.
Note, the coefficients $b_n$ are not individually determined
by the lattice input considered here, which mainly
constrains the gluonic vacuum sector.

Thus, the presently measured channels test the normalized-phase structure
without selecting a particular harmonic. The all-isoscalar modes
$\eta\eta\eta$ and $\eta\eta\eta'$ are different, since the triple-trace term
also contributes and makes them sensitive to the harmonic content of the
interaction~\cite{Fractional}. BESIII has already searched for
$X(2370)\to\eta\eta\eta'$ and reported an upper limit
\cite{BESetaetaetap}, making improved measurements of these channels
especially interesting.

A more complete treatment should eventually include glueball--quarkonium
mixing, scalar-mediated decays, and final-state interactions. The observed
$f_0(980)\eta'$ and $a_0(980)\pi$ contributions indicate that such effects
are phenomenologically relevant~\cite{BES2024,BES2026modes}.

\section{Outlook}

The resonance $X(2370)$ has the required quantum numbers, lies in the relevant mass
region and is now observed in several decay channels. The question is no
longer only whether a peak lies near a lattice mass: its decays can also
test the interaction used to describe it.

The ordinary determinant provides a natural starting point. Fractional
powers of its normalized phase connect this phenomenology to the
$\theta$ vacuum~\cite{Fractional}. Establishing a predominantly gluonic
interpretation still requires a combined understanding of production,
decays and mixing. But this resonance may teach us not only where the
glueball is, but also how the axial anomaly is realized in its hadronic
interactions.

\smallskip
\noindent\textit{Acknowledgements.} I thank my collaborators on the works discussed here, in particular
S.~Jafarzade,
Gy.~Kov\'acs, P.~Kov\'acs, R.~D.~Pisarski, F.~Rennecke, W.~Eshraim, S.~Janowski, and D.~H.~Rischke.
This contribution is based on my MESON2026 presentation and includes
subsequent developments.


\end{document}